\documentclass[prl,reprint,showpacs,superscriptaddress,showkeys,showpacs]{revtex4-1}%{revtex4}

\usepackage{amsmath,amsfonts,amssymb,amstext}
\usepackage[dvipdfm]{graphicx}
\usepackage[colorlinks,linkcolor=red]{hyperref}

\newcommand{\be}{\begin{equation}}
\newcommand{\ee}{\end{equation}}
\newcommand{\ben}{\begin{eqnarray}}
\newcommand{\een}{\end{eqnarray}}
\newcommand{\bes}{\begin{subequations}}
\newcommand{\ees}{\end{subequations}}
\newcommand{\bF}{\begin{figure}}
\newcommand{\eF}{\end{figure}}

\def\ket#1{ | #1 \rangle}

\def\tr{ {\rm{Tr }}\,}

\newcommand{\arxiv}[2][arxiv:]{\href{http://arxiv.org/abs/#1#2}{#1#2}}

\begin{document}
\title{Interpreting quantum discord through quantum state merging}
\date{\today}

\author{Vaibhav Madhok}
%\email{vmadhok@gmail.com}
\affiliation{Center for Quantum Information and Control, University of New Mexico, Albuquerque, NM 87131-0001, USA}

\author{Animesh Datta}
%\email{animesh.datta@physics.ox.ac.uk}
\affiliation{Clarendon Laboratory, Department of Physics, University of Oxford, OX1 3PU, United Kingdom}

\begin{abstract}
We present an operational interpretation of quantum discord based on the quantum state merging protocol. Quantum discord is the markup in the cost of quantum communication in the process of quantum state merging, if one discards relevant prior information. Our interpretation has an intuitive explanation based on the strong subadditivity of von Neumann entropy. We use our result to provide operational interpretations of other quantities like the local purity and quantum deficit. Finally, we discuss in brief some instances where our interpretation is valid in the single copy scenario.
\end{abstract}


\maketitle

%\section{Introduction}

Quantum information science is primarily aimed at harnessing the quantum structure of nature for information processing and computing tasks~\cite{nielsen00a}. This quest has met with considerable success over the last decade, but there has been substantial progress in the other direction as well. Information theory has provided a novel framework for unraveling the intricacies of quantum mechanics. Quantum correlations, as well as classical ones are now viewed as resources, whose interconvertibility is governed by quantum information theory~\cite{dhw08}. Foremost amongst these is evidently entanglement, which provides enhanced performance in several important tasks like communication, computation, metrology and others~\cite{pv07}.

In the realm of mixed-state quantum information, however, instances are known where quantum advantages are evidenced in the presence of little or no entanglement~\cite{dqc1}. Recently, quantum discord was proposed as the source behind this enhancement and first steps towards a formal proof have been taken~\cite{dsc08}. Quantum discord was originally suggested as a measure of quantumness of correlations~\cite{oz02}, and has since been studied in variety of systems and settings~\cite{belld,disc}. Initial motivation for its definition arose in the context of pointer states and environment induced decoherence~\cite{darwinism}. It has since been related to the performance to quantum and classical Maxwell's demons~\cite{demon}. Though satisfactory from a physical perspective, the benchmark for accepting some quantity as a resource in quantum information science is that it appear as the solution to an appropriate asymptotic information processing tasks. It is this \textit{operational} interpretation that has been lacking for quantum discord, and we now provide in this Letter. This also addresses a more fundamental dichotomy in quantum information science, where resources and their manipulations can have both thermodynamic and information theoretic interpretations independently, which are not intuitively or mathematically reconciled. Our Letter bridges this gap in the context of quantum discord, as well the quantum deficit and the local purity.

%\section{Quantum Discord}

Quantum discord aims at capturing all quantum correlations in a quantum state, including entanglement~\cite{oz02,hv01}. Quantum mutual information is generally taken to be the measure of total correlations, classical and quantum, in a quantum state. For two systems, $A$ and $B$, it is defined as $ I(A:B) = H(A) + H(B) -H(A,B).$ Here $H(\cdot)$ denotes the Shannon entropy of the appropriate distribution. For a classical probability distribution, Bayes' rule leads to an equivalent definition of the mutual information as $I(A:B) = H(A)-H(A|B).$  This motivates a definition of classical correlation in a quantum state. Suppose Alice and Bob share a quantum state $\rho_{AB} \in \mathcal{H}_A\otimes \mathcal{H}_B.$ If Bob performs the POVM  set $\{\Pi_i\},$ the resulting state is given by the shared ensemble $\{p_i,\rho_{A|i}\},$ where
$$
 \rho_{A|i} = \tr_{\!\!B}(\Pi_i\rho_{AB})/p_i,\;\;\;p_i=\tr_{\!\!A,B}(\Pi_i\rho_{AB}).
$$
A quantum analogue of the conditional entropy can then be defined as $\tilde{S}_{\{\Pi_i\}}(A|B)\equiv\sum_ip_iS(\rho_{A|i}),$ and an alternative version of the quantum mutual information can now be defined as $\mathcal{J}_{\{\Pi_i\}}(\rho_{AB}) = S(\rho_A)-\tilde{S}_{\{\Pi_i\}}(A|B),$ where $S(\cdot)$ denotes the von Neumann entropy of the relevant state. The above quantity depends on the chosen set of measurements $\{\Pi_i\}.$ To capture all the classical correlations present in $\rho_{AB},$ we maximize $\mathcal{J}_{\{\Pi_i\}}(\rho_{AB})$ over all $\{\Pi_i\},$ arriving at a measurement independent quantity
\be
\label{eq:J}
\mathcal{J}(\rho_{AB}) = \max_{\{\Pi_i\}}(S(\rho_A)-\tilde{S}_{\{\Pi_i\}}(A|B)).
\ee
Then, quantum discord is defined as~\cite{oz02}
\ben
\label{discexp}
\mathcal{D}(\rho_{AB}) &=& I(\rho_{AB})-\mathcal{J}(\rho_{AB}) \\
                 &=& S(\rho_A)-S(\rho_{AB})+\min_{\{\Pi_i\}}\tilde{S}_{\{\Pi_i\}}(A|B),\nonumber
\een
Since the conditional entropy is concave over the set of POVMs, which is convex, the minimum is attained on the extreme points of the set of POVMs, which are rank 1~\cite{dw04}. In the asymptotic limit, when Alice and Bob share $n$ copies of the state $\rho_{AB},$ we can define a regularized version of quantum discord as
\ben
\overline{\mathcal{D}}(\rho_{AB})&=&\lim_{n\rightarrow \infty} \frac{\mathcal{D}(\rho^{\otimes n}_{AB})}{n} \\ \nonumber
                                &\equiv & I(\rho_{AB}) - \overline{\mathcal{J}}(\rho_{AB}),
\een
where
\be
\overline{\mathcal{J}}(\rho_{AB}) = \lim_{n\rightarrow \infty} \frac{\mathcal{J}(\rho^{\otimes n}_{AB})}{n}.
\ee
The quantity $\overline{\mathcal{J}}(\rho_{AB})$ has an operational interpretation as a measure of classical correlations, as the distillable common randomness (DCR) with one-way classical communication~\cite{dw04}, which is identical to the regularized version of the measure of classical correlations as defined by Henderson and Vedral~\cite{hv01}. Whether there exists a `single-letter' expression for discord depends on its additivity, which is equivalent to that of the entanglement of formation since
$$
\overline{\mathcal{D}}(\rho_{AC}) = E_C(\rho_{AB}) + S(\rho_C) - S(\rho_{AC}),
$$
if $\rho_{ABC}$ is pure and $E_C(\cdot)$ is the entanglement cost, the regularized version of the entanglement of formation~\cite{pv07}. This can be obtained using the monogamy between DCR and $E_C$ \cite{kw04}. Following the counterexample to the additivity of the minimum output entropy\cite{h09} and therefore the entanglement of formation, we can conclude that quantum discord is not additive either. In fact, the subadditivity of minimum output entropy implies that in general, quantum discord is subadditive. Our endeavor here will be to provide an operational interpretation for quantum discord $\overline{\mathcal{D}}$ itself, without seeking recourse to its definition as the difference of total and classical correlations. To that end, we will employ the process of quantum state merging, which we describe next. For brevity, in the remainder of the paper, we will suppress explicit mention of the state $\rho_{AB}$ in the argument of quantities, and denote its von Neumann entropy as $S(A,B)$, its quantum discord when measurements are made on $B$ as $\mathcal{D}(A|B)$ etc.

%\section{Quantum state merging}

Consider a party Bob having access to some incomplete information $Y,$ and another party Alice having the missing the part $X.$  We can think of $X$ and $Y$ as random variables. If Bob wishes to learn $X$ fully, how much information must Alice send to him? Evidently, she can send $H(X)$ bits to satisfy Bob. However, Slepian and Wolf showed that she can do better, by merely sending $H(X|Y) = H(X,Y)-H(Y),$ the conditional information~\cite{thomascover}.  Since $H(X|Y) \leq H(X)$, Alice can take advantage of correlations between $X$ and $Y$ to reduce the communication cost needed to accomplish the given task. Quantum state merging protocol is the extension of the classical Slepian-Wolf protocol into the quantum domain where Alice and Bob share the quantum state $\rho_{AB}^{\otimes n}$, with each party having the marginal density operators $\rho_{A}^{\otimes n}$ and $\rho_{B}^{\otimes n}$ respectively. Let $\ket{\Psi_{ABC}}$ be a purification of $\rho_{AB}.$ We will assume later, without loss of generality, that Bob holds $C.$ The quantum state merging protocol quantifies the minimum amount of quantum information which Alice must send to Bob so that he ends up with a state close to $\ket{\Psi}_{B'BC}^{\otimes n},$ $B'$ being a register at Bob's end to store the qubits received from Alice. It was shown that in the limit of $n\rightarrow \infty$, and asymptotically vanishing errors, the answer is given by the quantum conditional entropy~\cite{how05,how07}: $S(A|B) = S(A,B) - S(B)$. When $S(A|B)$ is negative, Bob can obtain the full state almost perfectly with just local operations and classical communication. In addition, Alice and Bob can distill $-S(A|B)$ ebits which can be used to transfer additional quantum information in the future.

%We start with a version of the strong subadditivity of the von-Nuemann entropy. Suppose ABC is a composite quantum system.  Alice is in possession %of part A, of ABC, and Bob is in possession of parts B and C. Whenever we refer to system ABC henceforth, we will assume the the above scenario.
%
%
%\subsection{Quantum operations increase the cost of state merging}
%
%In this section we show that quantum operations on B  can never increase the cost of quantum communication needed by Alice in order to merge her state with B. However, such quantum operations can increase the cost of state merging. \AD{I am confused. I thought the amount of quantum communication needed by Alice in order to merge her state with B is the cost of state merging ? What did I miss ?} Consider again our composite system ABC. Assuming C to be Bob's ancilla, that starts in an initially pure state $\ket{0}.$ %Put some more stuff about strong sub additivity

A heuristic but intuitive argument for our interpretation of quantum discord beings with strong subadditivity. For a tripartite system, it states that~\cite{how07}
\be
\label{eq:ssa}
S(A|B,C) \leq S(A|B).
\ee
From the point of view of the state merging protocol, the above has a very clear interpretation: having more prior information makes state merging cheaper. Or in other words, throwing away information will make state merging more expensive. Thus, if Bob discards system C, it will increase the cost of quantum communication needed by Alice in order to merge her state with Bob. Our intent here shall be to relate this increase in cost of state merging to quantum discord between A and B.

%\subsection{Discord as a mark up in state merging}

In order to do so, we need to simulate an arbitrary quantum operation $\mathcal{E}$ (including measurements) on $B.$  For that, we assume $C$ to initially be in a pure state $\ket{\mathbf{0}}$, and a unitary interaction $U$ between $B$ and $C$. Letting primes denote the state of the system after $U$ has acted we have $S(A,B) = S(A,BC)$ as $C$ starts out in a product state with $AB$. We also have $I(A : BC) = I(A' : B'C')$. As discarding quantum systems cannot increase the mutual information, we get $I(A' : B') \leq I(A' : B'C')$. Now consider the state merging protocol between A and B in the presence of $C$. We have
\ben
S(A|B) &=& S(A) - I(A : B) \nonumber\\
&=& S(A) - I(A : BC) = S(A|BC).
\een
After the application of the unitary $U$, but before discarding the subsystem $C$, the cost of merging is still given by $S(A'|B'C') = S(A|B)$. In fact, this implies that one can always view the cost of merging state of system $A$ with $B$,  as the cost of merging $A$ with the system $BC$, where $C$ is some ancilla (initially in a pure state) with which $B$ interacts coherently through a unitary $U$. Such a scheme does not change the cost of state merging, as shown, but helps us in counting resources. Once we discard system $C$, we get
\be
\label{eq:Idiff}
I(A' : B') \leq I(A':B'C') = I(A:BC) = I(A : B),
\ee
or alternatively,
\be
\label{eq:Sdiff}
S(A'|B') \geq S(A'|B'C') = S(A|B).
\ee
If we now compute the marked up price in the state merging as we discard information, we recover $D = I(A:B) - I(A':B').$ We next show that the quantity $D$ is equal to quantum discord when our quantum operations are quantum measurements, and we seek to maximize $I(A' : B')$. Thus, discord is the minimum possible increase in the cost of quantum communication in order to perform state merging, when we perform a measurement on the party receiving the final state. This also addresses the asymmetry that is inherent is quantum discord. This is exhibited operationally in our interpretation since the state merging protocol is not invariant under exchanging the labels of the parties, as is the case for instance, in superdense coding, which provides an interpretation for quantum entanglement. We provide one later based on quantum discord for pure states.

%\subsection{Quantum discord and state merging}

We now show that $D$ reduces to quantum discord when we perform quantum measurements on $B$ and maximize $I(A':B').$ The state $\rho_{AB},$ under measurement of subsystem $B$, changes to $\rho_{AB} ' = \sum_{j} p_{j} \rho_{A|j} \otimes \pi_j,$ where $ \{\pi_j\} $ are orthogonal projectors resulting from a Neumark extension of the POVM elements. The unconditioned post measurement states of $A$ and $B$ are
$$
\rho_{A}' =\sum_{j} p_{j} \rho_{A|j} = \rho_{A},~~~\rho_{B}' = \sum_{j} p_{j} \pi_j.
$$ Computing the value of $I(A':B')$, we get
\ben
I(A':B') &=&  S(A') +S(B') - S(A',B'),\nonumber  \\
        &=& S(A') + H(p) -\big\{H(p) + \sum_{j} p_{j} S(\rho_{A|j})\big\}, \nonumber\\
            &=& S(A) - \sum_{j} p_{j} S(\rho_{A|j}).
\een
After maximization, it reduces to $\mathcal{J}(\rho_{AB})$, as in Eq. (\ref{eq:J}). The reduction to rank 1 POVMs follows as stated earlier.

We can also rewrite the expression for $D$ using Eq. (\ref{eq:Sdiff}) instead of Eq. (\ref{eq:Idiff}) as the increase of the conditional entropy $D =  S(A'|B') - S(A|B).$ The above expression makes our interpretation even more transparent. Quantum measurements on $B$ require us to discard quantum correlations between $A$ and $B$. This increases the average cost of quantum communication needed by $A$, to merge her post measurement state with $B$.  Since, $S(A'|B') = \sum_{j} p_{j} S( \rho_{A|j}) \geq S(A|B)$, there is always a mark up in the cost of state merging. Whatever information Alice and Bob loose through the measurement, results in making the quantum state merging more expensive by exactly the same amount. Note that our interpretation does not rely on the notion of entanglement. This is crucial since quantum discord is more general than entanglement, and most situations where discord is of interest, there is no entanglement~\cite{dqc1}. Hence an interpretation based on quantum entanglement would falter in those cases.

%\subsection{Properties of Discord}

We can now use our quantum state merging perspective to derive the various properties of discord. Since measurements on system $B$ will always result in either discarding of some information or at best preserving the original correlations, we will always get a price hike in state merging or at best we can hope to just break even. Hence, discord, which is the mark up, will always be greater than zero~\cite{oz02,datta10}.

Quantum discord of a state is zero if and only if the density matrix is block diagonal in its own eigenbasis, and the density matrix should be of the form $\rho_{AB} = \sum_{i} p_{i} \rho_{A|i} \otimes |\lambda_{i} \rangle \langle \lambda_{i} |,$ in the basis which diagonalizes $\rho_{B}$. If one makes the measurements on $B$, with projectors $| \lambda_{i} \rangle \langle \lambda_{i} |$, one gets $\rho^{M}_{AB} = \sum_{i} P_{j} \rho_{AB} P_{j} = \rho_{AB}.$ Thus, we have a choice of measurement which causes no loss of information, and thus we retain all the correlations between $A$ and $B$. Thus there is no mark up in the cost of merging a zero discord state.

The converse can be seen through the application of strong subadditivity in Eq.~(\ref{eq:ssa}). The equality of mutual information, $I(A:B)$, of the initial state and that of the state after quantum operations on $B$, $I(A':B') $ coincides with the equality condition for strong subadditivity. But this is exactly the condition for the nullity of quantum discord~\cite{datta10}. Thus a zero mark up in the cost of state merging implies zero discord.

An upper bound on discord is decided by an upper bound on the mark up we can get. Or equivalently, it is the upper bound on the information that can be lost due to a quantum operation on $B$. This is simply the entropy of the state at Bob's end, $S(B),$ since Bob cannot loose more information that there is at his disposal. Thus an upper bound  on discord is the von Neumann entropy of the measured subsystem.

%The nullity of quantum discord has implications to local broadcasting of quantum states~\cite{ls10,phh08}.
%\AD{Maybe we can say a few more things about this ?} Form our perspective, only when local measurements on $B$ does not disturb the overall state %(there is no leak of correlations to the environment), that we can relate these approaches. \AD{What does this last sentence mean ? You mean the %connection is there only for 0 discord states ?}

%\subsection{ Discord for pure states and connections with entanglement}

Finally, for pure states, quantum discord reduces to entanglement, and $S(A|B)$ = $S(A) - I(A:B)  = -S(A) \leq 0$. From our perspective, measurement destroys all the entanglement present between $A$ and $B$. Though the post measurement state merging of the state of $A$ with that of $B$ occurs at zero cost, they loose the $-S(A|B)$ potential Bell pairs, which could have been put to some use. Thus, entanglement gets a novel operational interpretation as the markup in merging a pure state, when $B$ is measured.

%\section{Other measures}

\textit{Other measures--} We now use our result to provide operational interpretations for a couple of other quantities that were introduced to capture the quantumness of correlations, with motivations different from those of discord. Since the entropy of a closed system cannot decrease, the total number of pure qubits in a closed system of state, measurement and observer cannot increase. Thus thermodynamically, the purity of quantum states is a resource, which needs quantification. The allowed set of operations in this paradigm are called closed local operations and classical communications (CLOCC), which is a modification of the local operations and the classical communications (LOCC) paradigm without free pure ancilla. The central task in this setup then is local purity distillation. If one-way communication is allowed from Bob to Alice, the rate for this task is given by~\cite{d05}
\be
\kappa(A|B)=\log(d_{AB}) -S(A,B) - \overline{\mathcal{D}}(A|B),
\ee
where $d_{AB} = \dim(\mathcal{H}_A\otimes \mathcal{H}_B).$ This immediately provides an operational interpretation for local purity.

Another measure of quantumness of correlations in the CLOCC framework is the quantum deficit, which is also thermodynamically motivated, and can be intuitively thought of as a form of nonlocality without entanglement, but with distinguishability~\cite{hhhossr05}. Like quantum discord, unlike entanglement, it can be nonzero for separable states. The corresponding measure of classical correlations, the classical deficit is known to be equivalent to the DCR~\cite{d05} in the asymptotic limit. So, the quantum deficit actually coincides with the quantum discord in this regime, and has the same operational interpretation as discord. For a finite number of copies, the quantum deficit is always a lower bound on the quantum discord~\cite{hhhossr05}, provided the measurements are restricted to von Neumann projections instead of POVMs, because free pure ancilla are not allowed. For a state of two qubits, this restriction collapses, since all rank 1 POVMs are indeed projectors. Finally, operational interpretations can easily be provided for other discord like measures, for instance, measurement induced disturbance (MID)~\cite{mid} using quantum state merging, by simple variations of our argument. There is however a caveat for two sided measures, since the properties of DCR with two-way communications are unknown, and the additivity of the corresponding measures is consequently open.

%\section{Discussion and Summary}

The end product of our information theoretic task is the regularized form of quantum discord. This was necessitated since the single-copy version of state merging does not lead to the conditional von Neumann entropy~\cite{berta09}. There are however, several interesting cases in which the rate of asymptotic state merging can be identified with the quantum discord of a single copy. Evidently, pure states are one such class, since in that case quantum discord reduces to entanglement. Since the DCR is additive for separable states~\cite{dw04}, we have a `single-letter' definition of discord for such states as well. A more interesting set of states for which discord is additive are the Bell-diagonal states, since their DCR is additive too~\cite{thld02}. Quantum discord of Bell diagonal states of two qubits is among the best understood~\cite{belld}, and we have now shown that this understanding can be exported to the asymptotic regime without further effort.

In conclusion, this Letter places quantum discord squarely in the midst of quantum informational concepts, and opens up the way for its manipulation as a resource in quantum information processing. By exhibiting deep connections between measures of correlations that arose out of varied motivations such as thermodynamics and the theory of open quantum systems, we hope that our work will serve as a stepping stone for a more comprehensive and unified understanding of quantum physics, thermodynamics and information theory.

%In the search for an operational interpretation of discord, one has to make use of the correlations which vanish because of measurements. Quantum state merging an ideal scenario to put these correlations to some use. Two states having the same value for $J$s but different $I$s will differ in how much cost there will be in merging the state of two parties. \AD{The last statement is certainly true, but does it imply anything more than its words ?}

AD thanks C. M. Caves, A. Shaji, K. Modi and G. Adesso for early discussions on an operational interpretation of quantum discord. This work was supported in part by the EPSRC (Grant No. EP/H03031X/1) and the EU~Integrated Project (QESSENCE). VM acknowledges the Center for Quantum Information and Control (CQuIC) where this work was done, and NSF Grant Nos. 0903953 and 0903692.

\textit{Note added} - After the completion of this study, during the writing of the present paper, another work appeared~\cite{cabmpw10} where an operational interpretation for quantum discord was provided using the entanglement consumption in an extended quantum state merging protocol.

%\bibliography{OpInterp}{}

\end{document}